# Sustained Robust Exciton Emission in Suspended Monolayer WSe₂ within the Low Carrier Density Regime for Quantum Emitter Applications

*Zheng-Zhe Chen[1,4,5], Chiao-Yun Chang[2], Ya-Ting Tsai[1,6], Po-Cheng Tsai[1,3], Shih-Yen Lin[1,3] and Min-Hsiung Shih[1,6,7]* *

[1]Research Center for Applied Sciences, Academia Sinica, Taipei 11529, Taiwan

[2]Department of Electrical Engineering, National Taiwan Ocean University, Keelung 20224, Taiwan

[3]Graduate Institute of Electronics Engineering, National Taiwan University, Taipei 10617, Taiwan

[4]Department of Physics, National Taiwan University, Taipei 10617, Taiwan

[5]Institute of Physics, Academia Sinica, Taipei 11529, Taiwan

[6]Department of Photonics and Institute of Electro-Optical Engineering, National Yang Ming Chiao Tung University, Hsinchu 30010, Taiwan

[7]Department of Photonics, National Sun Yat-sen University, Kaohsiung 80424, Taiwan

*E-mail address: mhshih@gate.sinica.edu.tw

**ABSTRACT**

The development of semiconductor optoelectronic devices is moving toward low power consumption and miniaturization, especially for high-efficiency quantum emitters. However, most of these quantum sources work at low carrier density region, where the Shockley-Read-Hall (SRH) recombination may dominant and seriously reduce the emission efficiency. In order to diminish the affection of carrier trapping and sustain a strong photoluminescence (PL) emission under low power pumping condition, we investigated on the influence of "Suspending" to monolayered tungsten diselenide ($WSe_2$), novel two-dimensional quantum material. Not only the PL intensity, but also the fundamental photoluminescence quantum yield (PLQY) has exhibited a huge, order-scale enhancement through suspending, even surprisingly, we found the PLQY improvement revealed far significantly under small pumping power and came out an exponential increase tendency toward even lower carrier density region. With its strong excitonic effect, suspended $WSe_2$ offers a solution to reduce carrier trapping and participate in non-radiative processes. Moreover, in the low-power range where SRH recombination dominates, suspended $WSe_2$ exhibited remarkably higher percentage of excitonic radiation compared to contacted $WSe_2$. Herein, we quantitatively demonstrate the significance of suspended $WSe_2$ monolayer at low carrier density region, highlighting its potential for developing compact, low-power quantum emitters in the future.

## INTRODUCTION

With the evolution of integrated optoelectronic systems and emerging display technologies, a new generation of semiconductor optoelectronic devices is being driven towards miniaturization, low power consumption, and high-speed modulation.[1-3] Recently, two-dimensional transition metal dichalcogenide (TMDC) semiconductor have attracted large attention due to their unique optical, electrical, thermal, and mechanical properties.[4-11] Most importantly, monolayer TMDCs possess strong-binding excitons due to its geometry confinement of crystal structure, thus become a promising candidate for developing compact and novel miniaturized light-emitting devices.[12-14] For example, the monolayer TMDC laser, exhibiting an extremely low threshold characteristic attributed to its strong excitonic effects, presents an opportunity for operating microscale lasers at significantly low excitation power levels.[15-19] Nowadays, monolayer TMDCs have been extensively studied and developed as plenty kinds of miniaturized optoelectronic devices, including light-emitting diodes (LEDs), lasers, optical modulators, solar cells, photodetectors, and field-effect transistor.[20-31]

Nowadays, developments for quantum and optical communication devices are undoubtedly the most critical issues among opto-electrical applications. TMDC material is also well-known to serve as a novel platform for compact quantum emitters, revealing even better performance or higher flexibility than the conventional quantum dots-based ones.[32-35] However, these single photon sources usually work at low-pumping power region, where the non-radiative Shockley-Read-Hall (SRH) recombination most seriously. And the growth of large-area monolayer TMDC materials through chemical vapor deposition (CVD) unavoidably leads to a high density of defects (e.g., S-vacancy).[36,37] These defects, particularly in the form of trap states, can lead to significant carrier trapping effects and subsequent reduction in luminescence efficiency, leading to challenges for TMDC to develop toward compact high-efficiency quantum applications in the upcoming generation.[38-40]

In addition, the interface effects of monolayer TMDCs are highly sensitive. When a monolayer TMDC semiconductor material is placed on a substrate, the interface effects may give rise to issues such as the induction of defect-mediated localized states or doping effects. These phenomena could potentially result in low efficiency of monolayer TMDCs light-emitting devices in the low-pumping power regime. Therefore, to as possible diminish the affection from interface effect, "Suspending" the monolayer TMDC may be an easy and valid solution.[41-43] Suspended monolayer TMDCs exhibit a significantly enhanced exciton effect, leading to a substantial increase in radiative emission efficiency.[44-46] While previous studies have reported the improvement in photoluminescence (PL) intensity of suspended TMDCs, the investigation of their individual influence on the SRH or Auger recombination rate remains unexplored. Particularly, in semiconductor light-emitting devices operating in the low-pumping power regime, the SRH recombination mechanism dominates.

In this study, we have compared the fundamental material characteristics differences of suspended and contacted monolayer tungsten diselenide ($WSe_2$). Through suspending the material, monolayer $WSe_2$ exhibited a dramatical photoluminescence quantum yield (PLQY) improvement, and showed an enhancement enlarge tendency toward smaller power pumping. Additionally, the values of SRH, radiative, and Auger recombination coefficients, revealing intrinsic changes in the carrier recombination pathways of suspended and contacted monolayer $WSe_2$ material have quantitatively been determined through time-resolved photoluminescence (TRPL) measurement. The SRH recombination rate in suspended $WSe_2$ is found to be significantly slower than that in the contacted sample, particularly under low-power pumping. The visualized evidence of the intrinsic recombination tendency change through suspending is demonstrated by low-temperature PL measurements, showing dramatically difference at low carrier density region and manifesting the potential of suspended monolayer $WSe_2$ for the future quantum optoelectronic devices.

**RESULTS**

To go under some deep investigations into material characteristics of suspended monolayer TMDCs, we first designed and etched down an μm-size square hole on the $SiN_x$ substrate, then the as-grown CVD-synthesized $WSe_2$ monolayer was transferred and covered on top of the square hole [Fig. 1(a)], forming suspended monolayer $WSe_2$. For all the measurements in this paper based on suspended and contacted $WSe_2$, the signal was collected at the red and blue mark site in Fig. 1(a), respectively. Two marks are separated more than 7 μm (the diameter of laser spot is around 1 μm) to avoid the interference from each site's photon emission or carrier diffusion. To confirm the existence of monolayer $WSe_2$, the Raman spectrum shows apparent $E_{2g}^1$, $A_{1g}$ peak at 251 cm$^{-1}$, 263 cm$^{-1}$, respectively, and with absence of interlayer $B_{2g}^1$ signal at near 310 cm$^{-1}$ [Fig. S1], which is consistent with the previous report.[47-49] Furthermore, based on the Raman spectrum analysis, we observe no significant peak shift between the suspended and contacted samples, indicating that there is no noticeable strain effect on our suspended $WSe_2$ devices.[50,51] In addition, it can be observed that the Raman peak intensities of the two structures are nearly comparable. In our measurement system, the input/output coupling efficiency of 2D materials on SiNx/Si structures and suspended structures results in only about a 5% optical enhancement. PL spectrum of contacted and suspended monolayer $WSe_2$ were presented in Fig. 1(b), which are obtained by pumping with 450-nm continuous-wave (CW) laser in room temperature. Under same pumping condition (pumping power: 2 μW), the suspended $WSe_2$ presented an order higher of PL intensity compared to the contacted one at near 745 nm, which comes from the excitonic emission of $WSe_2$. This phenomenon is due to the "strong exciton effect" on the suspended TMDC [Fig. 1(c)]. Although CVD-grown TMDC has its advantages to approach larger scale for practical devices, it still suffered from some defects' affection in the crystal lattice, e.g., S-vacancy. And these dense defect sites on the material surface often cause seriously trappings of carriers to localized state, going under defect-mediated localized state recombination, or so-called Shockley-Read-Hall (SRH) recombination. However, by suspending monolayer TMDC, the strengthened exciton effect can efficiently help carriers diminish trappings from localized states to go through the non-excitonic radiative SRH recombination. Therefore, the suspended $WSe_2$ will further demonstrate stronger excitonic radiation than the contacted one. Furthermore, as we looked into the full width at half maximum (FWHM) of normalized PL spectrum [Fig. S2], the suspended $WSe_2$'s spectrum has obvious smaller FWHM (55 meV) than the contacted one (65 meV). Also, the contacted $WSe_2$'s spectrum shows more intensity distribution at lower-energy part (1.5 ~ 1.65 eV), which stands for the emission from multiple-carrier (e.g., trion, biexciton) or defect-mediated recombination, which then becomes another evidence to the stronger exciton effect exhibited in suspended than in contacted $WSe_2$. It is worth to note that strongly PL enhancement from the suspended WSe2 had excluded out the factor of the optical interference in the vertical air suspended structure. However, the 200 nm shallow air regime would be enough to modify the carrier behaviors of the ultrathin TMDC atomic layer.

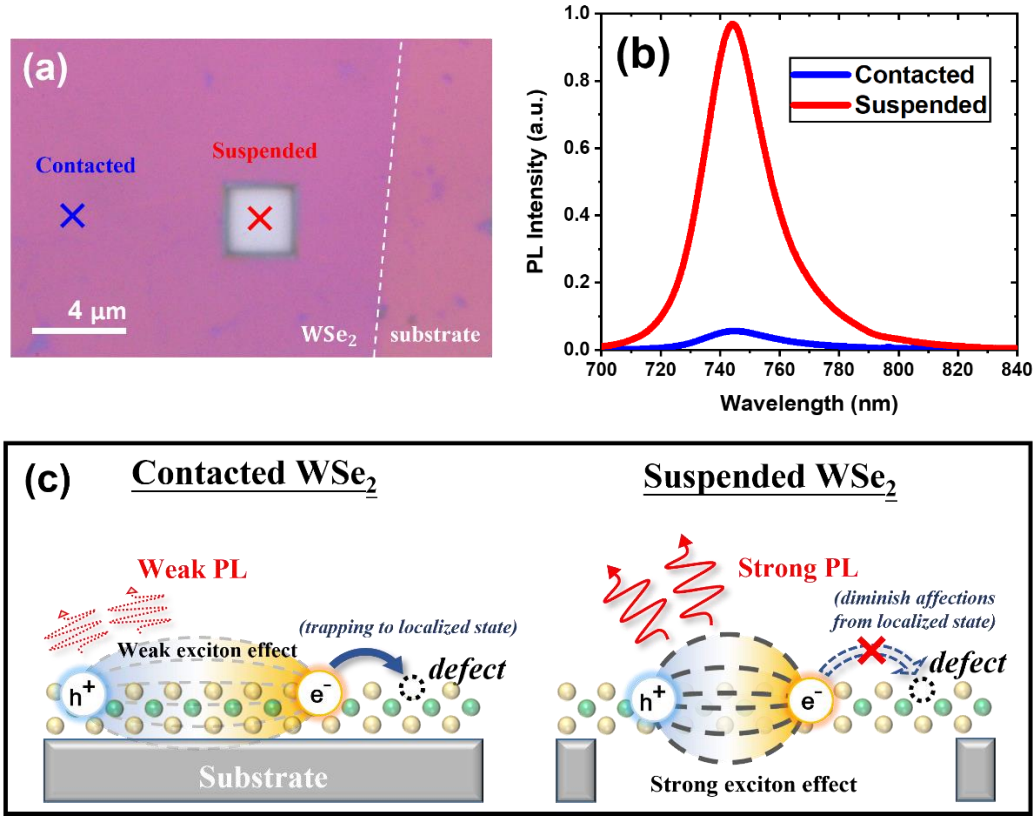

FIG. 1. (a) Optical image of monolayer WSe$_2$ suspended on a square hole with approximately 3 μm side length. The blue and red mark represent the pumping/collecting site of contacted and suspended WSe$_2$ in this paper, respectively. The white dash line is the edge of CVD-grown WSe$_2$ flake. (b) PL spectrum of contacted and suspended WSe$_2$ pumped by 450-nm CW laser in room temperature. (c) Schematic illustration of a comparison on how strong exciton effect in suspended WSe$_2$ can diminish the trapping from localized state and further reveals stronger PL intensity.

PL spectrum for both contacted and suspended WSe$_2$ under different pumping power have further been investigated, shown in Fig. 2(a) and Fig. 2(b), respectively. Both power-dependent PL spectrum have good positive correlation between its pumping power and intensity, however, the absolute intensity is much higher and FWHM is obviously narrower in the suspended WSe$_2$, which is coincided with the theory mentioned above. To quantify the difference of optical property and efficiency between contacted and suspended WSe$_2$, we integrated the PL intensity from full-wavelength range, and the total photon numbers can further be obtained after some calibration. Fig. 2(c) shows the total photon number collected from contacted and suspended WSe$_2$'s radiation per second under different pumping power. Being similar to the tendency of PL intensity, there is over an order more of photons generated from suspended WSe$_2$'s radiation per second ($\Phi_{photon}$) than the contacted WSe$_2$'s radiation under all different pumping power from 200 nW to 100 μW. To have an investigation into the WSe$_2$'s intrinsic change on quantum efficiency after suspending, the photoluminescence quantum yield (PLQY) has been calculated, which can be written as

$$PLQY\ (\%) = \frac{\Phi_{photon}\ (s^{-1})}{G.R._{pump}(s^{-1})} \times 100 \qquad (1)$$

where $\Phi_{photon}$ is the photon generated rate, $G.R._{pump}(s^{-1})$ represent the pumping generation rate whose value is related to the pumping source and measurement system (please see suppl. information S7). Fig. 2(d) shows the PLQY for contacted and suspended WSe$_2$ under different but a wide range of pumping generation rate per area $G.R._{pump}\ (cm^{-2}s^{-1})$. We found that the contacted WSe$_2$ (blue dots) exhibits no more than 1% of PLQY under any $G.R._{pump}$, which is consistent with the previous literature, in contrast, the suspended WSe$_2$ displays PLQY larger than 5%, and reached to about 12 % at its maximum.[52-54] With suspending the material, the PLQY can have over an-order of improvement, especially at low $G.R._{pump}$ region. By calculating the ratio between PLQY of contacted and suspended $WSe_2$, we surprisingly found the PLQY ratio reveals a dramatical increasement (fitted by exponential Chapman function) as lowering the $G.R._{pump}$ [Fig. 2(e)]. A 15-times PLQY enhancement exhibits at the lowest $G.R._{pump}$ in our experiment after suspending the material, giving the evidence of severe carrier trapping from defects in contacted $WSe_2$ within low carrier density region. Furthermore, in Fig. 2(d), it's not difficult to find that for both contacted and suspended $WSe_2$, the PLQY value does not have absolute positive or negative correlation to the $G.R._{pump}$, that is, the PLQY will slightly decrease as the pumping power is either too low or too high. That is probably due to the differences of carrier's distribution to distinct recombination pathway at different $G.R._{pump}$. To understand more details, we applied the ABC model which had been widely applied for semiconductor[55] and 2-D materials,[56,57]

$$G = An + Bn^2 + Cn^3 \qquad (2)$$

where G is the carrier generation/recombination rate at steady state. A, B, C represents the non-radiative SRH, radiative and Auger recombination coefficient, respectively. n is the carrier density. In the low carrier density region (small $G.R._{pump}$), most of the carrier may be trapped to localized state forming SRH recombination, which is nearly a non-radiative pathway at room temperature. As the carrier density increases, the more partial of carriers can form exciton and generate PL at the bang gap. While the number of carriers keeps increase (large $G.R._{pump}$), some non-radiative multi-carrier recombination, like Auger-like or exciton-exciton annihilation (EEA) recombination, may start to be dominant.[52,54,58,59] Looking back to Fig. 2(d), suspended $WSe_2$ is found to reach its maximal PLQY (red arrow) at $G.R._{pump} = 9 \times 10^{18}$ (cm$^{-2}$s$^{-1}$), however, contacted $WSe_2$ needs larger $G.R._{pump} = 5 \times 10^{19}$ (cm$^{-2}$s$^{-1}$) of pumping to get to its PLQY maximum (blue arrow). This is coincided with the theory that suspended $WSe_2$ exhibits stronger exciton effect.[42] When carrier concentration is not high, strong exciton effect in suspended $WSe_2$ can keep carriers from trapping to the localized state which makes it reveal larger radiative recombination coefficient B and smaller non-radiative SRH recombination coefficient A than the contacted $WSe_2$ (will be proved later). And that's also the reason why we can observe the largest PLQY enhancement at the small $G.R._{pumping}$ region.

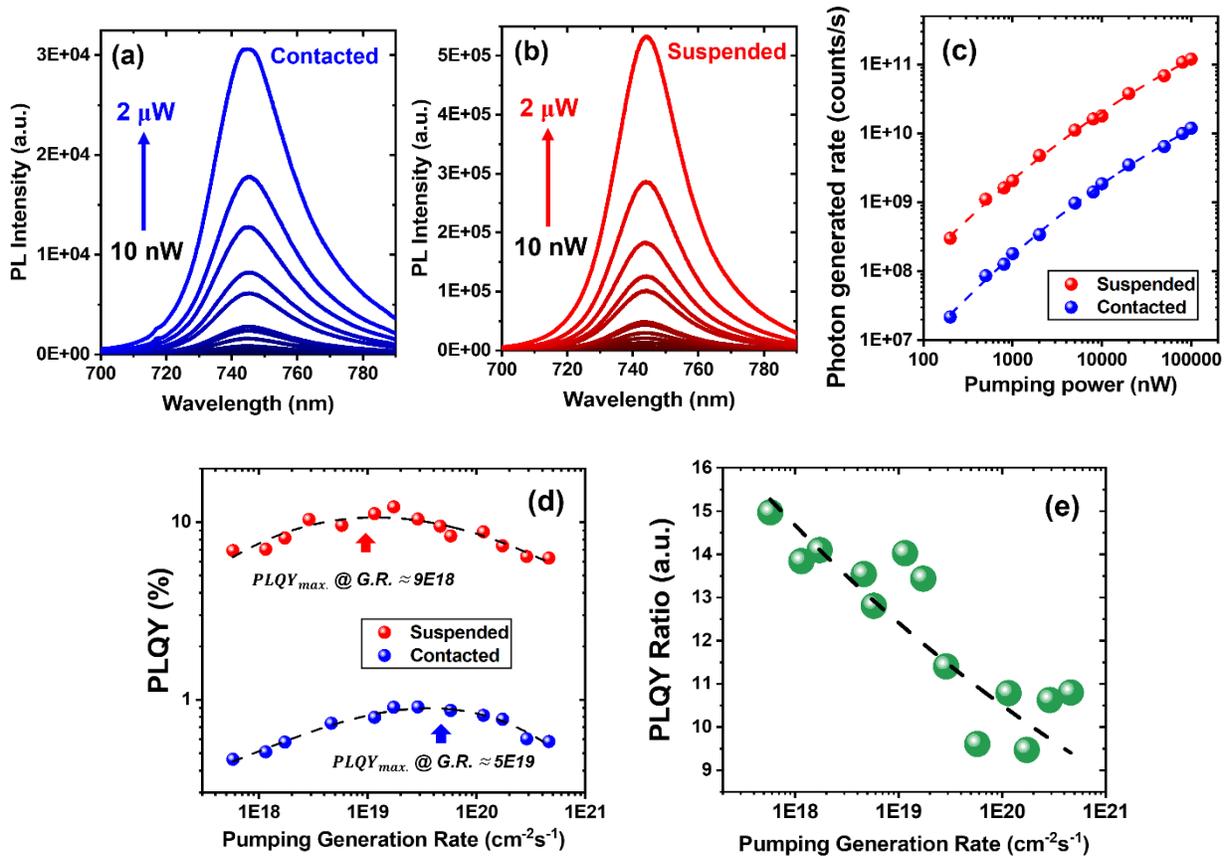

FIG. 2. PL spectrum with different pumping power (from 10 nW to 2 μW) of (a) contacted and (b) suspended WSe$_2$ in room temperature. (c) Photon generated rate on contacted (blue) and suspended (red) WSe$_2$ as a function of pumping power (ranging from 200 nW to 100 μW). (d) Plot of the PLQY with different pumping generation rate. (e) PLQY ratio between suspended and contacted WSe$_2$ in function of pumping generation rate. Dots represent the raw data from contacted (blue) and suspended (red) WSe$_2$. Dashed lines are polynomial fitting and exponential fitting for clearer tendency in (c), (d) and (e), respectively.

To have quantitative understanding to the carrier concentration and its recombination rate, time-resolved photoluminescence (TRPL) measurements were taken with our time-correlated single photon counting (TCSPC) system in room temperature (detail please see **Method**). And the measurements were centered to 745 nm, which is the exciton wavelength of $WSe_2$ in room temperature. Fig. 3(a), 3(b) shows the TRPL decay curve of contacted and suspended $WSe_2$, respectively, pumped by 450 nm 25 MHz-pulsed laser with average pumping power range from 0.2 to 5 μW. No matter in contacted or suspended case, the intensity exhibits a bi-exponential decay with time and the lifetime can be observed decreased with enlarging the pumping power, which is corresponded to the early theory and literature.[60,68] To obtain the recombination rate of radiative and non-radiative part, the TRPL intensity were then fitted by bi-exponential decay function,[60,61]

$$I\,(t) = A_r\,exp\,(\frac{-t}{\tau_r}) + A_{nr}\,exp\,(\frac{-t}{\tau_{nr}}) \qquad (3)$$

And averaged PL lifetime $\tau_{av}$ has further been calculated by

$$\tau_{av} = \frac{A_r\tau_r^2 + A_{nr}\tau_{nr}^2}{A_r\tau_r + A_{nr}\tau_{nr}} \qquad (4)$$

where t is the delay time. $A_r$, $A_{nr}$ represents the radiative and non-radiative decay coefficient, respectively. $\tau_r$, $\tau_{nr}$ represents the radiative and non-radiative lifetime, respectively. All the fitting results including $\tau_r$, $\tau_{nr}$ and $\tau_{av}$ are organized in Fig. S3(a). Averaged lifetime $\tau_{av}$ for contacted and suspended $WSe_2$ under different pumping power is plotted in Fig. 3(c). Here, we can clearly observe that the carrier lifetime of contacted $WSe_2$ is shorter in the low pumping power range. As the average pumping power increases, a crossover point occurs at near 1 μW of averaged pumping power. In the following discussion, we explore the physical significance of this phenomenon and further investigate the recombination mechanisms in suspended and contacted WSe$_2$ under different pumping power conditions by incorporating the recombination rate equation. Contacted WSe$_2$ have its $\tau_{av}$ shorter (faster recombination rate) than suspended WSe$_2$'s at low power region where the SRH recombination dominant, in contrast, suspended WSe$_2$ reveals its $\tau_{av}$ apparent shorter than contacted WSe$_2$'s above the intersection near 2 μW-averaged pumping power, where the radiative recombination starts to be dominant in the process. After reasonable bi-exponential fitting to TRPL decay curve, radiative (non-radiative) lifetime has been revealed, therefore the radiative (non-radiative) recombination rate can also be well-known [Fig. S3(b)-(e)]. Radiative and non-radiative recombination rate ($R_r$, $R_{nr}$) is the reciprocal of $\tau_r$ and $\tau_{nr}$, respectively.

$$R_{r(nr)} = \frac{1}{\tau_{r(nr)}} \qquad (5)$$

Shown in Fig. 3(d), suspended and contacted WSe$_2$ exhibits an intersection of exciton recombination rate (reciprocal of $\tau_{av}$) at near 1 μW of averaged pumping power. Suspended WSe$_2$ reveals apparent faster exciton recombination rate in higher pumping range, in which excitonic radiative recombination is dominant in the process, and present slower recombination rate under smaller pumping than contacted WSe$_2$, indicating the weaker tendency of suspended WSe$_2$ for defect-assisted SRH recombination, which is known for dominant in low-power region.

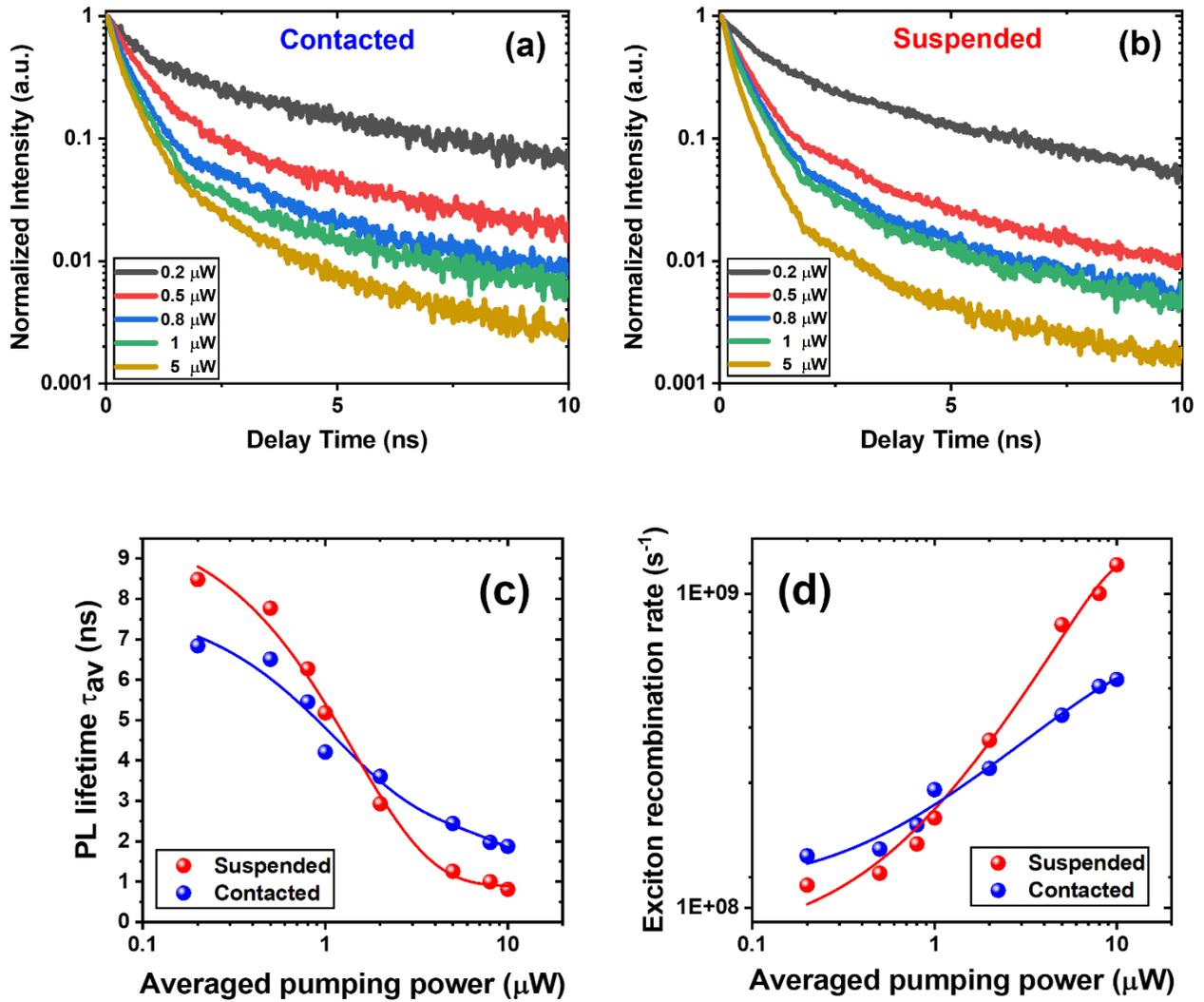

FIG. 3. Normalized TRPL decay curve of (a) contacted and (b) suspended WSe$_2$ with 450-nm pulsed laser of pumping in room temperature. (c) Plot of the average lifetime $\tau_{av}$ and (d) exciton recombination rate as a function of averaged pumping power. Dots represent the raw data and solid lines are biexponential fits to the data for clearer tendency.

From the results from TRPL measurement, we can calculate the carrier density n from each corresponded $G.R._{pump}$/pumping power.

At steady state,

$$\frac{dn}{dt} = G.R._{pump} - \frac{n}{\tau_{av}} = 0 \tag{6-1}$$

$$n = G.R._{pump} \cdot \tau_{av} \tag{6-2}$$

And to further comprehend the A, B, C recombination coefficient quantitively, we first start dig in the value of radiative recombination coefficient B by the relation,[56]

$$PLQY = \frac{Bn^2}{G.R._{pump}} \tag{7-1}$$

$$\Phi_{photon} \ (s^{-1}) \ = \ Bn^2 \tag{7-2}$$

Thus, by plotting the relation between total photon generated rate $\Phi_{photon} \ (s^{-1})$ and carrier density square $n^2 \ (cm^{-4})$, we can determine the slope as recombination coefficient B $(cm^4 s^{-1})$. For radiative recombination coefficient in contacted $WSe_2$ $(B_{con})$ equals to $9.24 \times 10^{-13} (cm^4 s^{-1})$, in contrast, suspended $WSe_2$ shows a $2.94 \times 10^{-11} (cm^4 s^{-1})$ value of coefficient $(B_{sus})$, which is over 30 times-larger than $B_{con}$ [Fig. 4(a)]. Next, coefficient A and C can also be determined by fitting to the G-n plot [Fig. 4(b)]. From eq. 2 (ABC model), we know the relation between carrier density n and the carrier generation/recombination rate G, thus the data from contacted (suspended) $WSe_2$ can further be fitted by the equation,[55,56]

$$G(s^{-1}) = An + B_{con(sus)}n^2 + Cn^3 \tag{8}$$

where the $B_{con}$ and $B_{sus}$ can be substituted with the value revealed in Fig. 4(a) to make the fitting easier and more accurate. The fitting results (dashed line in Fig. 4(b)) of SRH coefficient A and Auger coefficient C for contacted (suspended) WSe$_2$ were shown in table 1 (with determined coefficient B for comparison).

| Material | A (cm$^2$s$^{-1}$) | B (cm$^4$ s$^{-1}$) | C (cm$^6$ s$^{-1}$) |
|---|---|---|---|
| Contacted WSe$_2$ | 1.12 | $9.24 \times 10^{-13}$ | $2.07 \times 10^{-21}$ |
| Suspended WSe$_2$ | 0.52 | $2.94 \times 10^{-13}$ | $3.44 \times 10^{-21}$ |

Table 1. The fitting parameter of recombination rate equation for contacted and suspended WSe$_2$

According to the fitting results, it can be observed that the SRH coefficient A for contacted WSe$_2$ is approximately twice as large as that for suspended WSe$_2$. This finding provides evidence that the trend of defect-assisted non-radiative recombination is much stronger in contacted WSe$_2$ compared to suspended WSe$_2$. Moreover, to look into clearly on the distribution and tendency of carrier generation rate G, each process' recombination rate has been estimated based on their corresponded coefficient individually. The magnitude of recombination rate equals to $A \cdot n$ for SRH recombination rate ($R_{SHR}$), $B \cdot n^2$ for excitonic radiation recombination rate ($R_X$) and $C \cdot n^3$ for Auger recombination rate ($R_A$), respectively. Making a comparison to Fig. 4(c) and Fig. 4(d), $R_{SRH}$ is faster than $R_X$ at any carrier density (maybe will cross at higher carrier density) in contacted WSe$_2$, in contrast, suspended WSe$_2$ exhibits a reflection point at near $2 \times 10^{10}$ (cm$^{-2}$) of carrier density, where the $R_X$ starts to exceed $R_{SHR}$. Also, the percentage of $R_{SRH}$ among all recombination pathway have been estimated in Fig. S4. It's clear to realize that there exhibits about two times of $R_{SHR}$ percentage difference between contacted and suspended WSe$_2$ as the carrier density getting lower, which again prove the theory that strong exciton effect in suspended WSe$_2$ can effectively diminish the probability of carriers going through SRH recombination pathway. Nevertheless, both contacted and suspended WSe$_2$ suffered from a serious non-radiative multiple-carrier recombination process (e.g., Auger, EEA), leading to intrinsic low quantum efficiency of WSe$_2$'s radiation.[52-54] Based on the comprehensive discussion of TRPL and recombination rates, it can be concluded that for carrier injection densities below $1 \times 10^{10}$ (cm$^{-2}$), both suspended and contacted WSe$_2$ exhibit recombination rates predominantly governed by the SRH mechanism. The strong exciton effect in suspended WSe$_2$ effectively suppresses defect-assisted SRH recombination. Additionally, Fig. 2(d) illustrates a pronounced decrease in the PLQY of contacted WSe$_2$ when the pumping generation rates fall below $3 \times 10^{19}$ (cm$^{-2}$s$^{-1}$), with a significant drop of 70% relative to the maximum PLQY even at pumping generation rates below $1 \times 10^{18}$ (cm$^{-2}$s$^{-1}$). Conversely, the PLQY of suspended WSe$_2$ only decreases by 30% compared to the maximum PLQY at lower pumping powers. These results highlight the favorable characteristics of suspended WSe$_2$ for developing low-power devices, enabling high efficiency and ultra-low threshold laser applications in the field of optoelectronics.

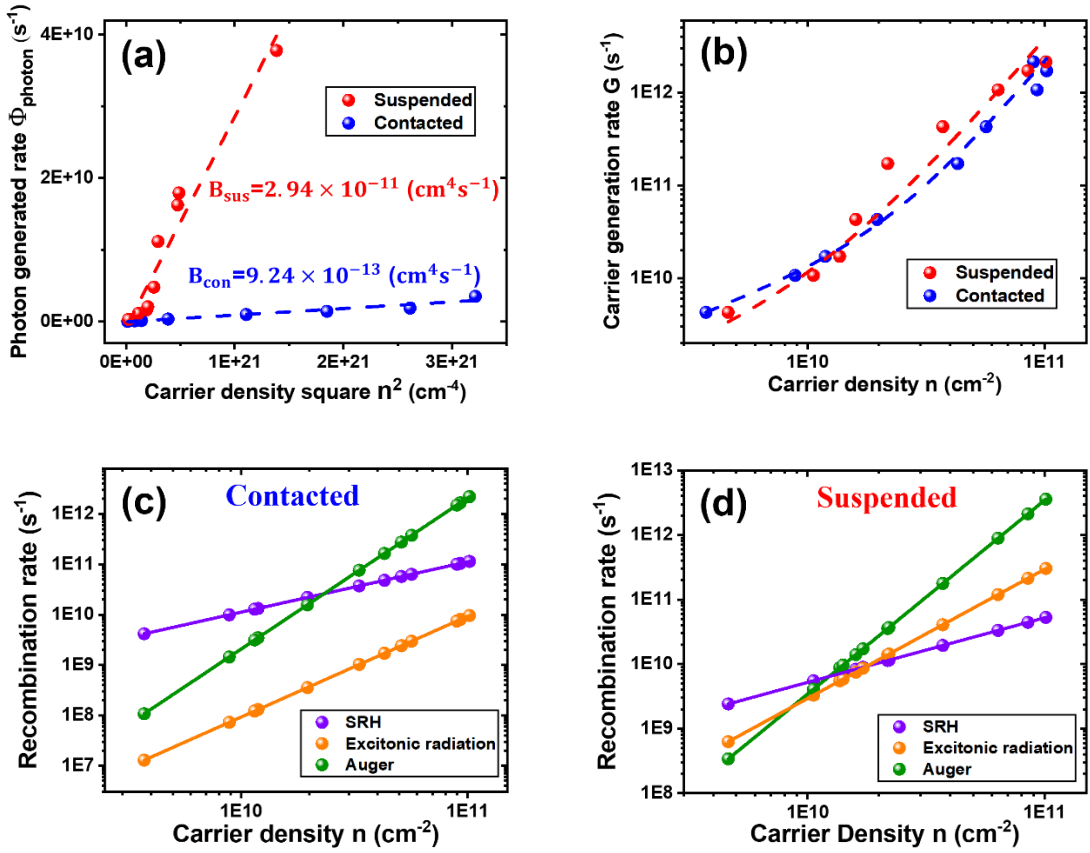

FIG. 4. (a) Plot of the photon generated rate $\Phi_{photon}$ as a function of carrier density square $n^2$. Dashed lines are linear fitting to the raw data to extract the slope as value of radiative recombination coefficient B. (b) Plot of the relation between carrier generation rate G and carrier density n. Dashed lines are fitting results from the ABC model (eq. 8). (c)(d) SRH (violet), exciton radiation (orange), Auger (green) recombination rate as a function of carrier density n in (c) contacted and (d) suspended $WSe_2$. Solid lines are polynomial fitting for clearer tendency.

For further realization to the defect-assisted localized-state radiation, the system was cooled down to liquid-nitrogen temperature (78 K) to diminish the interference from phonon scattering. Thus, WSe$_2$ PL spectrum become more distinct to identify the radiation from different quasi-particle/energy state in low temperature [Fig. 5(a), 5(b)]. Compared to PL spectrum in room temperature [Fig. S2], the spectrum in 78 K exhibit an apparent 50 meV blue-shift (from 1.67 to 1.72 eV) for both contacted and suspended WSe$_2$, which can be described using the Varshni's semiempirical equation.[52,56] And also, several distinguishable peaks from neutral exciton (X$^0$), negative-doped trion (X$^-$) and defect-mediated localized state (D) have shown at 1.744 eV, 1.716 eV and 1.631 eV under low temperature condition, respectively. The binding energy of trion and the energy difference between exciton and localized state emissions are estimated to be 28 meV and 113 meV, accordingly, which is corresponded to literature.[63,63,67] However, when we make a comparison between these two spectra, contacted WSe$_2$ brings out a quite more obvious shoulder at low energy, confirming the stronger tendency of localized state radiation in contacted WSe$_2$. To quantitatively analyze the localized state radiation discrepancy between contacted and suspended WSe$_2$, we further extracted distinct peaks from each pumping power's spectrum with typical gaussian fitting. Fig. S5 shows one of the spectra (contacted; 20 μW-pumping) fitted with three gaussian peaks, substantially representing defect-mediated localized state (green area), negative trion (orange area) and neutral exciton (blue area) from low to high energy. A little mismatch around 1.68 eV may come from the existences of some more complicated quasi-particle (e.g., biexciton, dark exciton),[65-67] which is not really clear enough to be identify here. Like Fig. S5 has been done, all the power-dependent spectrum from Fig. 5(a), 5(b) have been well-fitted in same way and the integrated PL intensity extracted from each state has been plotted in Fig. S6(a), 6(b). With increasing power (P), intensity (I) of localized state, trion and exciton increase exponentially, obeying the power law ($I \propto P^{\alpha}$). Fitting by the power law, contacted WSe₂ exhibits smaller constant $\alpha$ from exciton ($\alpha_{X^0} = 0.87$) than suspended WSe₂ does ($\alpha_{X^0} = 1.07$), again confirming the stronger exciton effect revealing in suspended WSe₂. To have an even clearer comparison, defect-mediated localized-state intensity percentage (LIP) has been calculated,

$$LIP~(\%) = \frac{A_D}{A_{X^0} + A_{X^-} + A_D} \times 100\% \tag{8}$$

where $A_{X^0}$, $A_{X^-}$ and $A_D$ represents the integrated PL intensity from exciton, trion and localized state, respectively. LIPs of contacted and suspended WSe₂ under different pumping power are plotted in Fig. 5(c). Since carriers in WSe₂ will be trapped in defect site more seriously in low carrier density region, LIP for contacted and suspended cases both increase when lowering the pumping power. The proportion of localized state intensity reveals over 50% higher in contacted than suspended WSe₂. Even a 60% of disparity is reached toward the lowest power pumping (0.1 μW) condition in our experiment, that is, LIP = 83% and 23% for contacted and suspended WSe₂, respectively, proving the concept that suspending can effectively diminish the affection of carrier trapping from localized state especially when the carrier density is low. Finally, in Fig. 5(d), a bar chart comparing excitonic (orange)/localized state (green) radiation percentage between contacted and suspended WSe₂ under 0.1 μW pumping is illustrated, showing the large disparity after suspending the material. Also, the mechanisms for both radiation procedures are depicted in the insets.

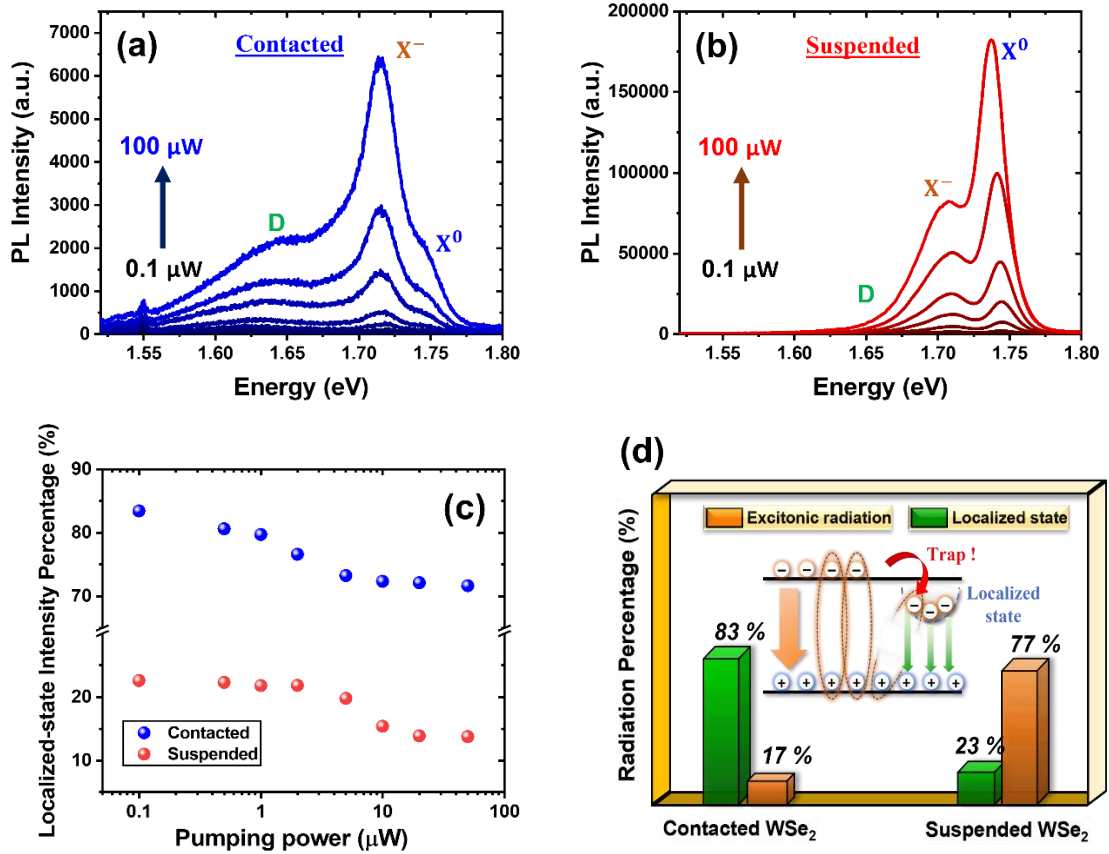

FIG. 5. PL spectrum with different pumping power (from 0.1 μW to 100 μW) of (a) contacted and (b) suspended WSe$_2$ in 78 K. X$^0$, X$^-$ and D stand for neutral exciton, negative trion and defect state radiation, respectively. (c) Plot of localized-state intensity percentage (LIP) in function of pumping power. (d) The bar chart of the radiation percentage comparison between contacted and suspended WSe$_2$ under the smallest pumping power (0.1 μW). Inset: Schematic of excitonic and localized state radiation mechanism.

**CONCLUSIONS**

In this study, we have demonstrated the superior performance of suspended WSe$_2$ as a compact light emitter compared to contacted WSe$_2$. At room temperature, suspended WSe$_2$ exhibited a PLQY of 12%, representing an improvement of over an order of magnitude compared to contacted WSe$_2$. By applying the ABC model for fitting the plot of carrier generation rate versus carrier density, we quantitatively determined the values of the Shockley-Read-Hall (SRH) coefficient A, radiative recombination coefficient B, and Auger coefficient C. The coefficient B in contacted WSe$_2$ (B$_{con}$) equals to $9.24 \times 10^{-13}(cm^4 s^{-1})$, in contrast, suspended WSe$_2$ shows a $2.94 \times 10^{-11}(cm^4 s^{-1})$ value of coefficient (B$_{sus}$), which is over 30 times-larger. On the other hand, the SRH coefficient A is approximately twice as large in the contacted sample, indicating a much stronger tendency for defect-assisted non-radiative recombination in contacted WSe$_2$. This observation is further supported by the visualization of low-temperature PL spectra. 60% of LIP disparity was discovered between contacted and suspended WSe$_2$ under the lowest pumping power. Our results showing the strong sustainability of exciton emission through suspending monolayer WSe$_2$ in the low carrier density regime have paved the way for the development of compact, low-power functional TMDC quantum light emitters in the future.

## SUPPLEMENTARY MATERIAL

The supplementary material contains additional Raman spectra, PL fitting results, and methods of device's fabrication process for the contacted/suspended WSe$_2$.

## ACKNOWLEDGEMENT

This work was supported by the Innovative Materials and Analytical Technology Exploration (i-MATE) program of Academia Sinica in Taiwan and the Ministry of Science and Technology (MOST) in Taiwan under Contract number MOST 108-2112-M-001-044-MY2, MOST 110-2112-M-001-053, and MOST 111-2112-M-001-078.

# REFERENCES

[1]X. Gan, D. Englund, D.V. Thourhout, and J. Zhao, Applied Physics Reviews **1**, 9(2), 021302 (2022).

[2]C.T. Phare, Y.-H. Daniel Lee, J. Cardenas, and M. Lipson, Nature Photonics **9**, 511 (2015).

[3]R. Yang, L. Zhou, H. Zhu, and J. Chen, Optics Express **23**, 28993 (2015).

[4]A. Splendiani, L. Sun, Y. Zhang, T. Li, J. Kim, C.-Y. Chim, G. Galli, and F. Wang, Nano Letters **10**, 1271 (2010).

[5]Q.H. Wang, K. Kalantar-Zadeh, A. Kis, J.N. Coleman, and M.S. Strano, Nature Nanotechnology **7**, 699 (2012).

[6]J.S. Ross, S. Wu, H. Yu, N.J. Ghimire, A.M. Jones, G. Aivazian, J. Yan, D.G. Mandrus, D. Xiao, W. Yao, and X. Xu, Nature Communications **4**, 1474 (2013).

[7]T. Shen, A.V. Penumatcha, and J. Appenzeller, ACS Nano **10**, 4712 (2016).

[8]P. Johari and V.B. Shenoy, ACS Nano **6**, 5449 (2012).

[9]G.-H. Lee, Y.-J. Yu, X. Cui, N. Petrone, C.-H. Lee, M.S. Choi, D.-Y. Lee, C. Lee, W.J. Yoo, K. Watanabe, T. Taniguchi, C. Nuckolls, P. Kim, and J. Hone, ACS Nano **7**, 7931 (2013).

[10]D. Jariwala, V.K. Sangwan, L.J. Lauhon, T.J. Marks, and M.C. Hersam, ACS Nano **8**, 1102 (2014).

[11]R. Rong, Y. Liu, X. Nie, W. Zhang, Z. Zhang, Y. Liu, and W. Guo, Advanced Science **10**, 2206191 (2023).

[12]J. Pu and T. Takenobu, Advanced Materials **30**, 1707627 (2018).

[13]F. Withers, O. Del Pozo-Zamudio, A. Mishchenko, A.P. Rooney, A. Gholinia, K. Watanabe, T. Taniguchi, S.J. Haigh, A.K. Geim, A.I. Tartakovskii, and K.S. Novoselov, Nature Materials **14**, 301 (2015).

[14]Y.-H. Chang, Y.-S. Lin, K. James Singh, H.-T. Lin, C.-Y. Chang, Z.-Z. Chen, Y.-W. Zhang, S.-Y. Lin, H.-C. Kuo, and M.-H. Shih, Nanoscale **15**, 1347 (2023).

[15]J. Shang, C. Cong, Z. Wang, N. Peimyoo, L. Wu, C. Zou, Y. Chen, X.Y. Chin, J. Wang, C. Soci, W. Huang, and T. Yu, Nature Communications **8**, 543, (2017).

[16]S. Wu, S. Buckley, J.R. Schaibley, L. Feng, J. Yan, D.G. Mandrus, F. Hatami, W. Yao, J. Vučković, A. Majumdar, and X. Xu, Nature **520**, 69 (2015).

[17]L. Zhao, Q. Shang, Y. Gao, J. Shi, Z. Liu, J. Chen, Y. Mi, P. Yang, Z. Zhang, W. Du, M. Hong, Y. Liang, J. Xie, X. Hu, B. Peng, J. Leng, X. Liu, Y. Zhao, Y. Zhang, and Q. Zhang, ACS Nano **12**, 9390 (2018).

[18]Y. Li, J. Zhang, D. Huang, H. Sun, F. Fan, J. Feng, Z. Wang, and C.Z. Ning, Nature Nanotechnology **12**, 987 (2017).

[19]X. Ge, M. Minkov, S. Fan, X. Li, and W. Zhou, Npj 2D Materials and Applications **3**,16 (2019).

[20]C. Wang, F. Yang, and Y. Gao, Nanoscale Advances **2**, 4323 (2020).

[21]C.-H. Lee, G.-H. Lee, A.M. Van Der Zande, W. Chen, Y. Li, M. Han, X. Cui, G. Arefe, C. Nuckolls, T.F. Heinz, J. Guo, J. Hone, and P. Kim, Nature Nanotechnology **9**, 676 (2014).

[22]D.-H. Lien, M. Amani, S.B. Desai, G.H. Ahn, K. Han, J.-H. He, J.W. Ager, M.C. Wu, and A. Javey, Nature Communications **9**, 1229 (2018).

[23]J. Gu, B. Chakraborty, M. Khatoniar, and V.M. Menon, Nature Nanotechnology **14,** 1024 (2019).

[24]D. Andrzejewski, H. Myja, M. Heuken, A. Grundmann, H. Kalisch, A. Vescan, T. Kümmell, and G. Bacher, ACS Photonics **6**, 1832 (2019).

[25]J.S. Ross, P. Klement, A.M. Jones, N.J. Ghimire, J. Yan, D.G. Mandrus, T. Taniguchi, K. Watanabe, K. Kitamura, W. Yao, D.H. Cobden, and X. Xu, Nature Nanotechnology **9**, 268 (2014).

[26]Y. Ye, Z.J. Wong, X. Lu, X. Ni, H. Zhu, X. Chen, Y. Wang, and X. Zhang, Nature Photonics **9**, 733 (2015).

[27]O. Salehzadeh, M. Djavid, N.H. Tran, I. Shih, and Z. Mi, Nano Letters **15**, 5302 (2015).

[28]E.Y. Paik, L. Zhang, G.W. Burg, R. Gogna, E. Tutuc, and H. Deng, Nature **576**, 80 (2019).

[29]H. Lin, C. Chang, C. Yu, A.B. Lee, S. Gu, L. Lu, Y. Zhang, S. Lin, W. Chang, S. Chang, and M. Shih, Advanced Optical Materials **10**, 2200799 (2022).

[30]H.-M. Li, D.-Y. Lee, M.S. Choi, D. Qu, X. Liu, C.-H. Ra, and W.J. Yoo, Scientific Reports **4**, (2014).

[31]H. Wang, L. Yu, L.-H. Lee, Y. Shi, A. Hsu, M.L. Chin, L.-J. Li, M. Dubey, J. Kong, and T. Palacios, Nano Letters **12**, 4674 (2012).

[32]T. Gao, M. Von Helversen, C. Antón-Solanas, C. Schneider, and T. Heindel, Npj 2D Materials and Applications **7**, 4 (2023).

[33]S. Guo, S. Germanis, T. Taniguchi, K. Watanabe, F. Withers, and I.J. Luxmoore, ACS Photonics (2023).

[34]S. Cianci, E. Blundo, F. Tuzi, G. Pettinari, K. Olkowska-Pucko, E. Parmenopoulou, D.B.L. Peeters, A. Miriametro, T. Taniguchi, K. Watanabe, A. Babinski, M.R. Molas, M. Felici, and A. Polimeni, Advanced Optical Materials **11**, 2202953 (2023).

[35]J.-P. So, K.-Y. Jeong, J.M. Lee, K.-H. Kim, S.-J. Lee, W. Huh, H.-R. Kim, J.-H. Choi, J.M. Kim, Y.S. Kim, C.-H. Lee, S. Nam, and H.-G. Park, Nano Letters **21**, 1546 (2021).

[36]J. Hong, Z. Hu, M. Probert, K. Li, D. Lv, X. Yang, L. Gu, N. Mao, Q. Feng, L. Xie, J. Zhang, D. Wu, Z. Zhang, C. Jin, W. Ji, X. Zhang, J. Yuan, and Z. Zhang, Nature Communications **6**, 6293 (2015).

[37]A.A. Koós, P. Vancsó, M. Szendrő, G. Dobrik, D. Antognini Silva, Z.I. Popov, P.B. Sorokin, L. Henrard, C. Hwang, L.P. Bíró, and L. Tapasztó, The Journal of Physical Chemistry C **123**, 24855 (2019).

[38]Z. Chi, H. Chen, Z. Chen, Q. Zhao, H. Chen, and Y.-X. Weng, ACS Nano **12**, 8961 (2018).

[39]P. Taank, R. Karmakar, and K.V. Adarsh, Surface and Interface Analysis (2023).

[40]F. Zhong, J. Ye, T. He, L. Zhang, Z. Wang, Q. Li, B. Han, P. Wang, P. Wu, Y. Yu, J. Guo, Z. Zhang, M. Peng, T. Xu, X. Ge, Y. Wang, H. Wang, M. Zubair, X. Zhou, P. Gao, Z. Fan, and W. Hu, Small **17**, 2102855 (2021).

[41]H. Shi, R. Yan, S. Bertolazzi, J. Brivio, B. Gao, A. Kis, D. Jena, H.G. Xing, and L. Huang, ACS Nano **7**, 1072 (2013).

[42]M. Drüppel, T. Deilmann, P. Krüger, and M. Rohlfing, Nature Communications **8**, 2117 (2017).

[43]D. Lloyd, X. Liu, J.W. Christopher, L. Cantley, A. Wadehra, B.L. Kim, B.B. Goldberg, A.K. Swan, and J.S. Bunch, Nano Letters **16**, 5836 (2016).

[44]Y. Yu, Y. Yu, C. Xu, Y.-Q. Cai, L. Su, Y. Zhang, Y.-W. Zhang, K. Gundogdu, and L. Cao, Advanced Functional Materials **26**, 4733 (2016).

[45]X. Ao, X. Xu, J. Dong, and S. He, ACS Applied Materials & Interfaces **10**, 34817 (2018).

[46]X. Zhang, D. Sun, Y. Li, G.-H. Lee, X. Cui, D. Chenet, Y. You, T.F. Heinz, and J.C. Hone, ACS Applied Materials & Interfaces **7**, 25923 (2015).

[47]P. Tonndorf, R. Schmidt, P. Böttger, X. Zhang, J. Börner, A. Liebig, M. Albrecht, C. Kloc, O. Gordan, D. Zahn, S. Michaelis de Vasconcellos, and R. Bratschitsch, Optics Express **21**, 4908 (2013).

[48]H. Terrones, E.D. Corro, S. Feng, J.M. Poumirol, D. Rhodes, D. Smirnov, N.R. Pradhan, Z. Lin, M.A.T. Nguyen, A.L. Elías, T.E. Mallouk, L. Balicas, M.A. Pimenta, and M. Terrones, Scientific Reports **4**, 4215 (2014).

[49]Y. Pan and D.R.T. Zahn, Nanomaterials **12**, 3949 (2022).

[50]I. Niehues, R. Schmidt, M. Drüppel, P. Marauhn, D. Christiansen, M. Selig, G. Berghäuser, D. Wigger, R. Schneider, L. Braasch, R. Koch, A. Castellanos-Gomez, T. Kuhn, A. Knorr, E. Malic, M. Rohlfing, S. Michaelis De Vasconcellos, and R. Bratschitsch, Nano Letters **18**, 1751 (2018).

[51]S.B. Desai, G. Seol, J.S. Kang, H. Fang, C. Battaglia, R. Kapadia, J.W. Ager, J. Guo, and A. Javey, Nano Letters **14**, 4592 (2014).

[52]D.-H. Lien, S. Z. Uddin, M. Yeh, M. Amani, H. Kim, J. W. Ager III, E. Yablonovitch, A. Javey, Science **364**, 468 (2019).

[53]Y. Lee, J.D.S. Forte, A. Chaves, A. Kumar, T.T. Tran, Y. Kim, S. Roy, T. Taniguchi, K. Watanabe, A. Chernikov, J.I. Jang, T. Low, and J. Kim, Nature Communications **12**, 7095 (2021).

[54]Y. Lee, T.T. Tran, Y. Kim, S. Roy, T. Taniguchi, K. Watanabe, J.I. Jang, and J. Kim, ACS Photonics **9**, 873 (2022).

[55]S. L. Chuang, Physics of Photonic Devices, Wiley (2009).

[56]O. Salehzadeh, N.H. Tran, X. Liu, I. Shih, and Z. Mi, Nano Letters **14**, 4125 (2014).

[57]Xumin Bao, Yuejun Liu, Guoen Weng, Xiaobo Hu, and Shaoqiang Chen, Quantum Electronics **48**, 7 (2018).

[58]S.Z. Uddin, N. Higashitarumizu, H. Kim, J. Yi, X. Zhang, D. Chrzan, and A. Javey, ACS Nano **16**, 8005 (2022).

[59]Y. Lee, G. Ghimire, S. Roy, Y. Kim, C. Seo, A.K. Sood, J.I. Jang, and J. Kim, ACS Photonics **5**, 2904 (2018).

[60]J. Wang, J. Ardelean, Y. Bai, A. Steinhoff, M. Florian, F. Jahnke, X. Xu, M. Kira, J. Hone, X.-Y. Zhu, Sci. Adv. **5**, eaax0145 (2019).

[61]H. Liu, T. Wang, C. Wang, D. Liu, and J. Luo, The Journal of Physical Chemistry C **123**, 10087 (2019).

[62]Varshni, Y., Physica, Amsterdam **34**, 149 (1967)

[63]S. Golovynskyi, O.I. Datsenko, D. Dong, Y. Lin, I. Irfan, B. Li, D. Lin, and J. Qu, The Journal of Physical Chemistry C **125**, 17806 (2021).

[64]J. Huang, T.B. Hoang, and M.H. Mikkelsen, Scientific Reports **6**, 22414 (2016).

[65]Y. You, X.-X. Zhang, T.C. Berkelbach, M.S. Hybertsen, D.R. Reichman, and T.F. Heinz, Nature Physics **11**, 477 (2015).

[66]Z. Li, T. Wang, Z. Lu, C. Jin, Y. Chen, Y. Meng, Z. Lian, T. Taniguchi, K. Watanabe, S. Zhang, D. Smirnov, and S.-F. Shi, Nature Communications **9**,3719 (2018).

[67]T.P. Lyons, S. Dufferwiel, M. Brooks, F. Withers, T. Taniguchi, K. Watanabe, K.S. Novoselov, G. Burkard, and A.I. Tartakovskii, Nature Communications **10**, 2330 (2019).

[68]C. Choi, J. Huang, H.-C. Cheng, H. Kim, A.K. Vinod, S.-H. Bae, V.O. Özçelik, R. Grassi, J. Chae, S.-W. Huang, X. Duan, K. Kaasbjerg, T. Low, and C.W. Wong, Npj 2D Materials and Applications 2, (2018).

# S1. Raman spectrum of the monolayer contacted and suspended WSe₂

Raman spectrum of (a) contacted and (b) suspended WSe₂ measured in this paper were presented in Figure S1. The characteristic $E_{2g}^1$, $A_{1g}$ peak at 251 cm⁻¹, 263 cm⁻¹ represent in-plane and out-of-plane vibration of WSe₂ structure, respectively, which is consistent to previous reports.[1-3] And the absence of interlayer $B_{2g}^1$ signal at near 310 cm⁻¹ confirms the monolayer structure. On the other hand, from the comparison of Figure S1a and Figure S1b, no clear Raman peak shift was found, showing neglectable strain affection to the suspended WSe₂. Additionally, the relative intensity Raman spectrum of the contacted and suspended monolayer WSe₂ is also presented in FIG. S1 (c). Comparing the peak intensities of the $E_{2g}^1$ in these two structures reveals a difference of only about 5%.

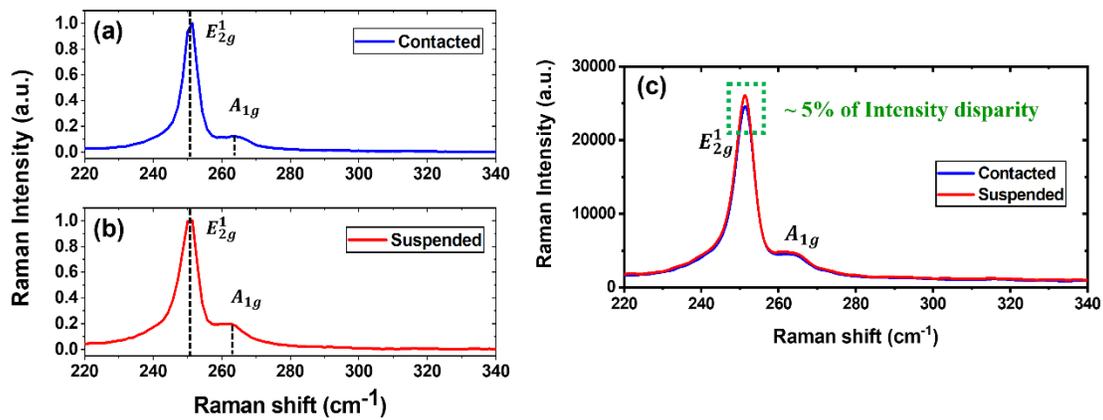

FIG. S1. Normalized intensity of the Raman spectrum for (a) the contacted and (b) suspended monolayer WSe₂. (c) Relative intensity Raman spectrum of the contacted and suspended monolayer WSe₂.

**S2. Normalized PL spectrum from contacted and suspended WSe₂**

Normalized PL spectrum from Figure 1b is revealed in Figure S2 with x-axis change to energy (eV) scale. There is an apparent FWHM difference between contacted (65 meV) and suspended (55 meV) WSe₂. Broaden of contacted WSe₂'s spectrum at left side of peak (low energy) may refer to defect-assisted SRH recombination.

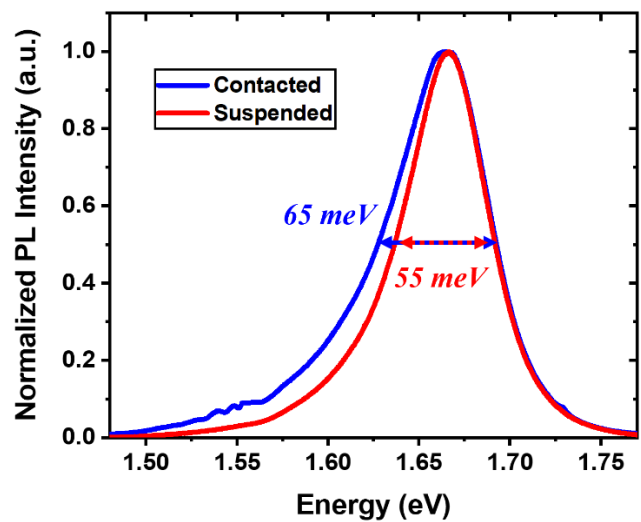

FIG. S2. Normalized PL spectrum of contacted and suspended WSe₂ under 2 μW of pumping power.

## S3. Radiative and non-radiative lifetime extracted from TRPL

Values of radiative lifetime $(\tau_r)$, non-radiative lifetime $(\tau_{nr})$ and averaged lifetime $(\tau_{av})$ extracted from different pumping power have been organized in Figure S3a, and further plotted in Figure S3b, Figure S3c and Figure 3c in a function of pumping power, respectively. Suspended WSe$_2$ reveals apparently shorter $\tau_r$ in the pumping range where radiative recombination dominant (Figure S3b). In contrast, suspended WSe$_2$ exhibits relatively longer $\tau_{nr}$ under any pumping power than contacted WSe$_2$ (Figure S3c).

## (a)

| Averaged Pumping Power (μW) | Contacted | | | Suspended | | |
|---|---|---|---|---|---|---|
| | $\tau_{nr}$ | $\tau_r$ | $\tau_{av}$ | $\tau_{nr}$ | $\tau_r$ | $\tau_{av}$ |
| **0.2** | 0.901 | 8.673 | 6.841 | 0.997 | 8.664 | 8.482 |
| **0.5** | 0.734 | 8.113 | 6.507 | 0.811 | 8.123 | 7.772 |
| **0.8** | 0.617 | 7.438 | 5.450 | 0.721 | 7.121 | 6.268 |
| **1** | 0.504 | 7.021 | 4.211 | 0.668 | 6.781 | 5.178 |
| **2** | 0.396 | 6.428 | 3.604 | 0.512 | 5.361 | 2.929 |
| **5** | 0.336 | 5.321 | 2.440 | 0.425 | 2.960 | 1.256 |
| **8** | 0.311 | 5.073 | 1.972 | 0.387 | 2.668 | 0.998 |
| **10** | 0.291 | 4.820 | 1.875 | 0.373 | 2.300 | 0.809 |

**(unit: ns)**

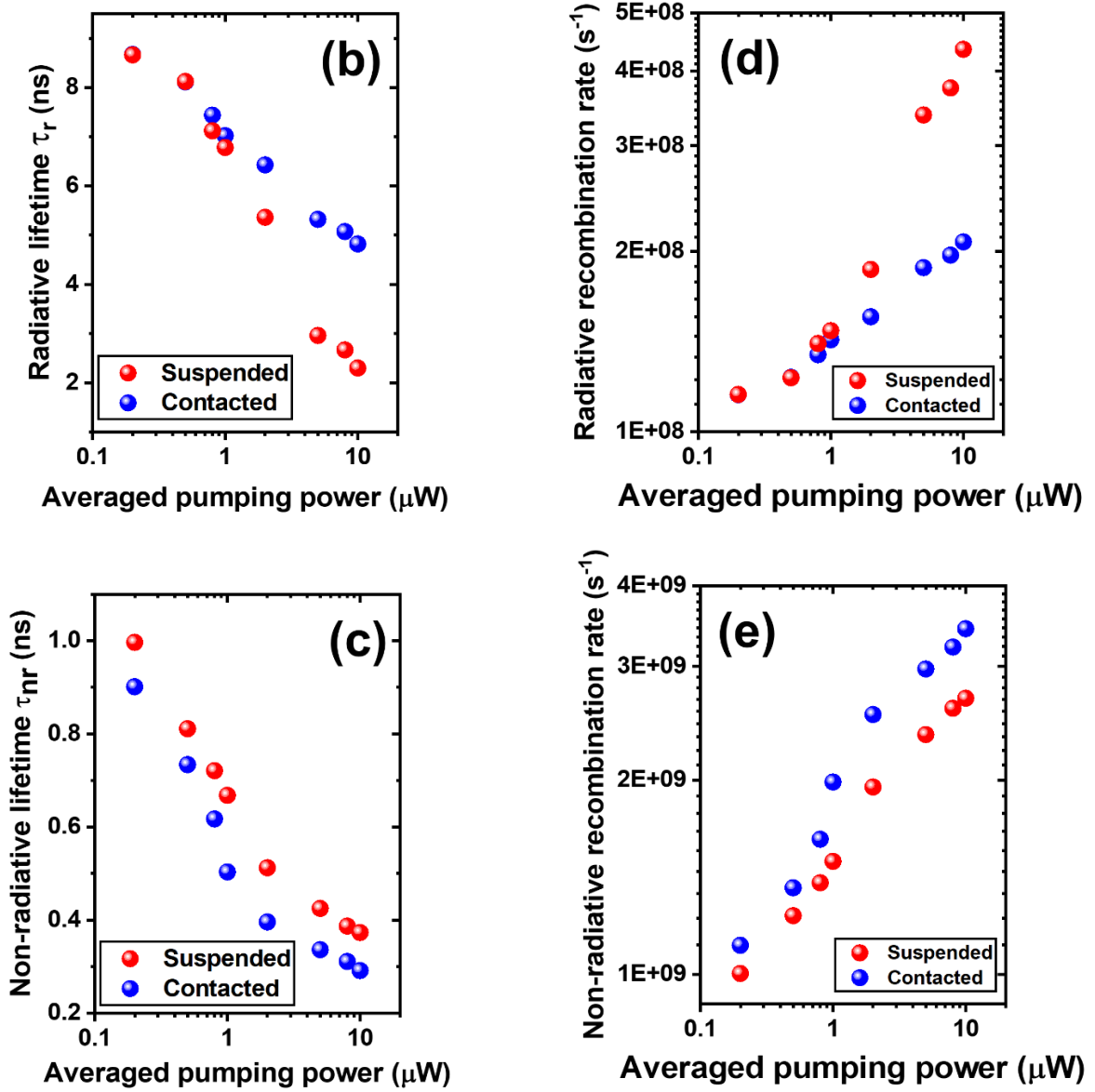

FIG. S3. (a) List of all fitted $\tau_r$, $\tau_{nr}$ and $\tau_{av}$ results (pumping power range from 200 nW to 100 μW). Plot of (b) radiative lifetime $\tau_r$ and (c) non-radiative lifetime $\tau_{nr}$ in a function of pumping power.

## S4. Percentage of SRH recombination rate

To quantitatively clarify the affection of SRH recombination in all carrier recombination pathway in contacted and suspended $WSe_2$. $R_{SRH}$ percentage is calculated by

$$R_{SRH} \; percentage \; (\%) = \frac{R_{SRH}}{R_{SRH} + R_X + R_A} \times 100\%$$

and plotted in Figure S4 in a function of carrier density. It's clear to see that $R_{SRH}$ shows a larger percentage as the carrier density decreases. And also, the most apparent difference between contacted and suspended $WSe_2$ happens at the low carrier density region.

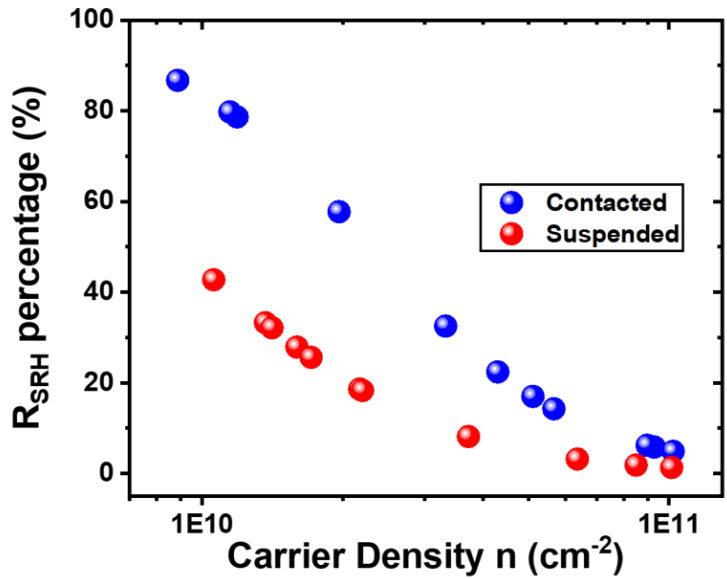

FIG. S4. Plot of percentage of $R_{SRH}$ among all recombination rate in a function of carrier density.

**S5. Integrated PL intensity of Exciton, Trion and Defect state in 78 K**

Under liquid-nitrogen temperature (78 K), the interference from the phonon scattering can be diminished, which makes the radiation of each state can be easier identified from the PL spectrum. Like example in Figure S5, with gaussian-peak fitting to the PL spectrum, the radiation from exciton $X^0$ (blue area), trion $X^-$ (orange area) and defect-mediated localized state D (green area) can be distinguished, and the intensity from each state can also be integrated individually.

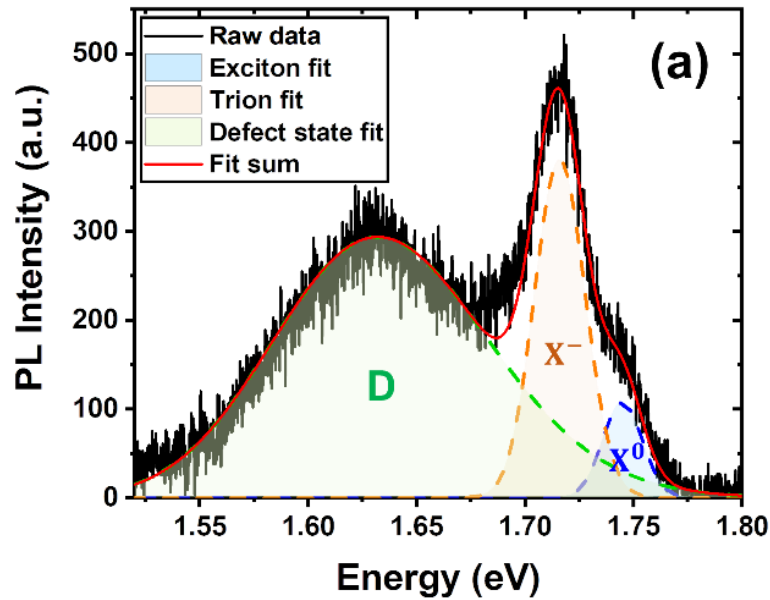

FIG. S5. PL spectrum of contacted WSe$_2$ under 20 μW pumping fitted with three gaussian peaks, representing exciton $X^0$ (blue area), trion $X^-$ (orange area) and defect-mediated localized state D (green area).

## S6. Integrated PL intensity of Exciton, Trion and Defect state in 78 K

With gaussian-peak fitting to the PL spectrum measured in 78 K, the integrated intensity from exciton, trion and defect state can be identified/calculated individually. Exhibited in Figure S6a and Figure S6b, radiation intensity either from $X^0$, $X^-$ or D follows the power law ($I \propto P^\alpha$). For defect state radiation, contacted WSe$_2$ shows slightly larger $\alpha_D$ (= 0.89) than suspended WSe$_2$ (=0.82). Suspended WSe$_2$, in contrast, reveals larger $\alpha_{X^0}$ (=1.07) and $\alpha_{X^-}$ (=1.05) than contacted WSe$_2$ ($\alpha_{X^0}$ =0.87; $\alpha_{X^-}$ =1.04), proving stronger exciton effect exists in suspended WSe$_2$.

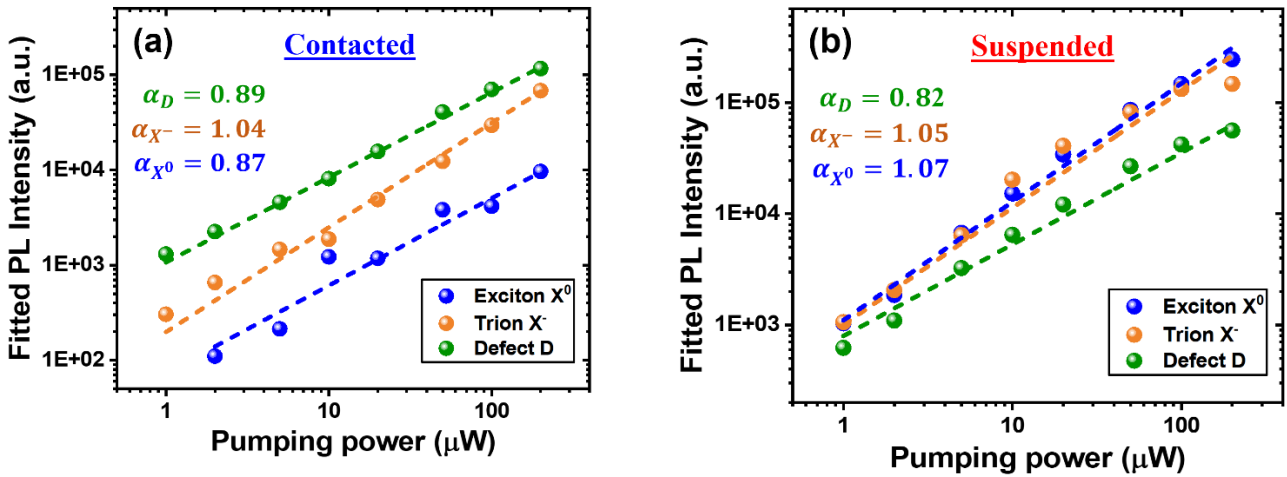

FIG. S6. Plot of (a) contacted and (b) suspended WSe$_2$'s integrated PL intensity from exciton $X^0$ (blue), trion $X^-$ (orange) and defect-mediated localized state D (green) in a function of pumping power in 78 K. Solid dots represent the raw data, while the dashed line represents the fitting of power law ($I \propto P^\alpha$) to these raw data.

## S7. Calculation of pumping generation rate $\text{G.R.}_{pump}$ $(cm^{-2}s^{-1})$

Pumping generation rate can be calculated by formula,[4]

$$\text{G.R.}_{pump} \ (cm^{-2}s^{-1}) = \frac{P \cdot \alpha \cdot (1-R)}{A \cdot h\nu}$$

where P is the pumping power (W), $\alpha$ is the absorption coefficient, R is the Fresnel reflection coefficient, A represents the laser spot area $(cm^{-2})$ and $h\nu$ is the energy of pumping source (450 nm CW laser) (J). In this study, we imply $\alpha$ = 0.01, R = 0.2 for monolayer WSe$_2$. And excitation laser spot size in our system is about $0.5 \times 0.5 \times \pi$ (μm$^2$) under 100x N.A. = 0.55 objective lens. We assume absorption coefficient $\alpha$, reflection coefficient R and laser spot size A are independent to the pumping power, thus, $\text{G.R.}_{pump}$ is proportional to the magnitude of pumping power.

.

## Methods

### Device Fabrication

First, a 200-thick $Si_3N_4$ layer was deposited on a silicon substrate through low pressure chemical vapor deposition (LPCVD). Next, to fabricate a square hole with reasonable size (larger than laser spot size, but not too large in case of the influence from break or strain), we use electron beam lithography (EBL) to pattern the pattern. The electron beam resist PMMA was spin-coated on $Si_3N_4$ for EBL development. After development, the substrate then went under an inductively coupled plasma reactive-ion etching (ICP-RIE) process. The residual resist PMMA was removed with oxygen plasma in the same ICP-RIE. Finally, after the square hole has been characterized, CVD-grown monolayer $WSe_2$ was further wet transferred on top of the $Si_3N_4$ layer, and make sure fully covered the square hole, forming uniform suspended $WSe_2$.[5]

### Optical Measurement

All the spectrum were measured by the home-build confocal-microscope μ-PL system. The steady-state PL spectrum were pumped by a 450-nm wavelength CW diode laser focused by a 100×, N.A.= 0.55 objective lens with spot size around 1 μm. And the PL radiation is collected by the same objective lens and further recorded/analyzed by a spectrograph with thermoelectrically cooled charge-coupled device detector. The resolution of spectrum was approximately 0.06 nm. The Raman spectrum were taken under same condition, but with a 532-nm wavelength CW laser of excitation instead. For TRPL measurement, a 450-nm wavelength pulsed diode laser with a 25-MHz repetition rate and 500-ps pulse width was used as the excitation source. Signals from TRPL were then collected using a single photon avalanche diode connected with a time-corrected single photon counting (TCSPC) module. For low temperature measurements, the sample was placed in a cryostat chamber, evacuated to about $2 \times 10^{-6}$ torr and cooled down by liquid nitrogen. Other pumping and measured system were same as in room temperature.

## References

[1]P. Tonndorf, R. Schmidt, P. Böttger, X. Zhang, J. Börner, A. Liebig, M. Albrecht, C. Kloc, O. Gordan, D. Zahn, S. Michaelis de Vasconcellos, and R. Bratschitsch, Optics Express **21**, 4908 (2013).

[2]H. Terrones, E.D. Corro, S. Feng, J.M. Poumirol, D. Rhodes, D. Smirnov, N.R. Pradhan, Z. Lin, M.A.T. Nguyen, A.L. Elías, T.E. Mallouk, L. Balicas, M.A. Pimenta, and M. Terrones, Scientific Reports **4**, 4215 (2014).

[3]Y. Pan and D.R.T. Zahn, Nanomaterials **12**, 3949 (2022).

[4]O. Salehzadeh, N.H. Tran, X. Liu, I. Shih, and Z. Mi, Nano Letters **14**, 4125 (2014).

[5]J.-K. Huang, J. Pu, C.-L. Hsu, M.-H. Chiu, Z.-Y. Juang, Y.-H. Chang, W.-H. Chang, Y. Iwasa, T. Takenobu, and L.-J. Li, ACS Nano **8**, 923 (2014)